\documentclass[AMA,STIX1COL]{WileyNJD-v2}
\articletype{Article Type}%

\received{26 April 2016}
\revised{6 June 2016}
\accepted{6 June 2016}

\usepackage{listings}
\usepackage{amsmath}
\usepackage{graphicx}
\usepackage{caption}
\usepackage{amssymb}
\usepackage{multirow}
\usepackage{algorithm}
\usepackage{subfig}
\usepackage{hhline}
\usepackage{tabularx}
\usepackage{amssymb}

\newcommand\tomasz[1]{#1}

\newcommand\Fawzy[1]{#1}

\begin{document}

\title{IoTSim-Osmosis-RES: Towards autonomic renewable energy-aware osmotic computing}

\author[1]{Tomasz Szydlo*}
\author[1]{Amadeusz Szabala}
\author[1]{Nazar Kordiumov}
\author[1]{Konrad Siuzdak}
\author[1]{Lukasz Wolski}
\author[2]{Khaled Alwasel}
\author[3,4]{Fawzy Habeeb}
\author[4]{Rajiv Ranjan}

\authormark{Tomasz Szydlo \textsc{et al}}

\address[1]{\orgdiv{Institute of Computer Science}, \orgname{AGH University of Science and Technology}, \orgaddress{\state{Al. Mickiewicza 30, 30-059 Krakow}, \country{Poland}}}

\address[2]{\orgdiv{College of Computing and Informatics}, \orgname{Saudi Electronic University}, \orgaddress{\state{Riyadh}, \country{Saudi Arabia}}}

\address[3]{\orgdiv{College of Computer Science}, \orgname{University of Jeddah}, \orgaddress{\state{Jeddah}, \country{Saudi Arabia}}}

\address[4]{\orgdiv{The School of Computing}, \orgname{Newcastle University}, \orgaddress{\state{Newcastle Upon Tyne}, \country{United Kingdom}}}

\corres{*\email{tszydlo@agh.edu.pl},
{kalwasel@gmail.com},
{f.m.m.habeeb2@newcastle.ac.uk},
{raj.ranjan@ncl.ac.uk}}

%\presentaddress{This is sample for present address text this is sample for present address text}

\abstract[Summary]{Internet of Things systems exists in various areas of our everyday life. For example, sensors installed in smart cities and homes are processed in edge and cloud computing centres providing several benefits that improve our lives. The place of data processing is related to the required system response times - processing data closer to its source results in a shorter system response time. The Osmotic Computing concept enables flexible deployment of data processing services and their possible movement, just like particles in the osmosis phenomenon move between regions of different densities. At the same time, the impact of complex computer architecture on the environment is increasingly being compensated by the use of renewable and low-carbon energy sources. However, the uncertainty of supplying green energy makes the management of Osmotic Computing demanding, and therefore their autonomy is desirable. In the paper, we present a framework enabling osmotic computing simulation based on renewable energy sources and autonomic osmotic agents, allowing the analysis of distributed management algorithms. We discuss the challenges posed to the framework and analyze various management algorithms for cooperating osmotic agents. \tomasz{In the evaluation we show that changing the adaptation logic of the osmotic agents, it is possible to increase the self-consumption of renewable energy sources or increase the usage of low emission ones.}}


%\jnlcitation{\cname{%
%\author{Williams K.}, 
%\author{B. Hoskins}, 
%\author{R. Lee}, 
%\author{G. Masato}, and 
%\author{T. Woollings}} (\cyear{2016}), 
%\ctitle{A regime analysis of Atlantic winter jet variability applied to evaluate HadGEM3-GC2}, %\cjournal{Q.J.R. Meteorol. Soc.}, \cvol{2017;00:1--6}.}

\maketitle

%\footnotetext{\textbf{Abbreviations:} ANA, anti-nuclear antibodies; APC, antigen-presenting cells; IRF, interferon regulatory factor}

%%%%%%%%%%%%%%%%%%%%%%%%%%%%%%%%%%%%%%%%%%%%%%%%%%%%%%%%%%%%%%%%%%%%%%%

%% main text
\section{Introduction}

%The goal is to create a module that will include new energy-aware placement algorithms, extendible abstractions for various renewable energy sources and energy management policies. There is no simulator with this kind of functionality - that’s one of the facts which is motivating us. The next fact is that the European Union is going to achieve climatic neutrality in 2050. What more, the surplus of produced energy is mostly sent to the grid instead of using it for computing. We want to reach this goal and solve the problems above by extending one of the existing frameworks: IoTSim-Osmosis. \par

%This is a simulator destined to test and validate osmotic computing applications by entering an infrastructure configuration and simulated application logics on resources-level and getting results of each IoT transaction and osmotic application, battery and power consumption on IoT devices, edge, cloud and SD-WAN infrastructures (Software Defined Wireless Area Network). It is built over the CloudSim engine - authors created a different package for Osmosis core implementation. Our plan is to do something very similar like earlier - our module seems to be a set of classes which will be used by the original IoTSim-Osmosis framework (placement depending on green energy production).  \par

The development of Internet of Things (IoT) systems covering smart homes, telemedicine, and the fourth industrial revolution is possible due to the key technology enablers. Wireless communication, cloud computing, edge data centres and increased networks throughput enable efficient sensor data processing. However, this makes managing the operations of IoT systems more complex. Hence, the promising programming paradigm for such IoT systems is Osmotic Computing. Like in the osmosis phenomenon the processing of data can be moved between devices in a continuum of cloud computing, edge computing and IoT devices. As a result, it is possible to enhance the current properties and performance of IoT ecosystems, such as shortening the time of data processing, reducing the cost of maintenance, increasing reliability or reducing the amount of data sent to clouds. 

The development of computing infrastructure for modern edge computing IoT systems is associated with an increased demand for energy. Several solutions are introduced to reduce the emission of pollutants due to the useage of traditional energy sources, such as more economical processors and computer architectures. However, the adoption of renewable energy for data processing seems to be a promising approach. Still, it requires intelligent management of the infrastructure and the data stream processing flows. Moreover, given the dynamic execution context of IoT ecosystems and the uncertain nature of renewable energy sources, IoT ecosystems would become a non-trivial challenge in production environments. As such, new solutions and prototypes are required to increase renewable energy usage, productivity and profitability in IoT osmotic contexts.

%Moreover, given the dynamic execution context of devices in IoT systems resulting from the uncertainty of renewable energy sources, their usage in production environments can be a challenge. 

%However, existing simulators for IoT systems mainly focus on how data is processed and, to a certain extent, are limited to external factors influencing their operation. 

This paper proposes a new simulation model that covers a set of functionalities in the context of sustainable and autonomic IoT ecosystems. As well, we extended the \textit{IoTSim-Osmosis}\cite{iotsimosmosis} simulator and added our proposed model, which enable research on sustainable and autonomic IoT systems. The concept of osmotic computing allows for flexible flow migration between edge and cloud computing centres and osmotic microelements. Information about the availability of renewable sources can be reflected in the osmosis process itself - i.e. by the migration of processing to use green energy as efficiently as possible. Given the complexity of the adaptation process due to the size of IoT systems, this mechanism should be self-adaptive. A promising approach to the realization of this concept is the use of \textit{osmotic agents}\cite{osmotic_agents} that follows Autonomic Computing concept \cite{ac}. Cooperating agents associated with devices and computing centres can exchange messages and adapt to the available energy and the desired properties of the system. For example, in environmental monitoring systems, osmotic computing should ensure data provision and processing continuity during natural disasters. In a stable situation, provide the greatest possible use of renewable energy. 

The rest of the paper is organized as follows. Section 2 describes the background of the research. Section 3 discusses the energy modelling used in the simulator, while section 4 describes the implementation details. Section 5 provides the evaluation details for several adaptation algorithms. Section 6 analyzes the related work and compares their functionalities. Finally, the paper is concluded in section 6.
%The introduced mechanisms - modelling of renewable energy sources and osmotic agents - allows for the research on management algorithms for osmotic computing.

\subsection{Challenges}
Research on energy-aware osmotic computing and realization of the simulation environment requires facing the following challenges: 

\begin{itemize}
    \item \textit{use of historical data} - due to the high uncertainty of weather conditions – e.g. wind speed and solar radiation level, the efficient analysis of adaptive osmotic computing management algorithms should be based on historical data regarding renewable energy sources. \tomasz{It is discussed in section 4.1.}
    
    \item \textit{decentralized management} - the scale of IoT systems makes centralized management of process migration and data flow control processes usually inefficient. Therefore, it is essential to introduce mechanisms enabling the implementation of distributed management algorithms. Furthermore, it will allow reacting to changes in the environment in which the devices operate. \tomasz{We are discussing the Autonomic Computing concept and its realization in section 2.4 and 4.2 respectively}.

    \item \textit{adaptation mechanisms} - reacting to changes in the availability of renewable energy in osmotic computing may include the reorganization of the data stream processing flows. Other mechanisms may include changing the routing of data streams between different data processing centres or limiting sensor data streams by, e.g. reducing the accuracy of sensor data reading or decreasing the reading frequency. \tomasz{The discussion regarding introduced mechanisms is in section 4.3.}
\end{itemize}

\subsection{Contribution}
\Fawzy{To solve the above challenges, this paper proposes a novel simulation model that encompasses a number of features in the context of self-sustaining and autonomous IoT ecosystems. The following are the paper's key contributions:}

\begin{itemize}
    \item \Fawzy{we extended the IoTSim-Osmosis simulator with our novel model that consists of a set of functionalities in the context of sustainable and autonomic IoT ecosystems,}

    \item \Fawzy{we evaluated IoTSim-Osmosis-RES using a case study that focused on managing renewable energy sources in IoT systems. The simulation results summarise IoTSim-Osmosis-RES functionality for assessing various parameters such as the level of solar radiation, usage of the RES, usage of the low emission sources, and the IoT device battery capacity.}
\end{itemize}

\section{Background}
\subsection{Osmotic computing}
Modern Internet of Things systems are usually based on a programming paradigm that emphasizes data flows and stream data processing\cite{szydlo2017}. A characteristic property of this class of systems is that they are composed of computational processes analyzing data streams and passing the results between them. Furthermore, the processes are deployed on the available hardware infrastructure, i.e. computational clouds, edge datacenters and devices themselves. The concept that generalizes this approach is Osmotic Computing\cite{osmosis} which provides elasticity to such systems enabling processes to move between devices similarly as solvent molecules through the osmotic membrane towards the computational clouds or edge datacenters. Moreover, it allows for the fulfilment of desired system requirements such as time of processing or energy efficiency by adapting to the current system operational context.

In osmotic computing, an application is defined as a graph composed of Microelements (MEL)\cite{mel}. They represent specific functionalities and can be deployed on cloud and edge resources. MELs can migrate within a \textit{software-defined membrane} which is an abstraction of a virtual environment spread over the available hardware infrastructure - edge and cloud resources. The IoT application in this concept is modelled as \textit{Osmotic Flow}\cite{flow} and described as a directed graph representing the transformation of data streams from IoT devices by processes defined in the graph nodes. The Osmotic Computing concept enables the deployment and management of stream transformation tasks within the available computing resources at the edge of the network and in the computing cloud.

\subsection{Renewable Energy Sources}
\begin{figure}[t]
\centering
\includegraphics[width=0.9\textwidth]{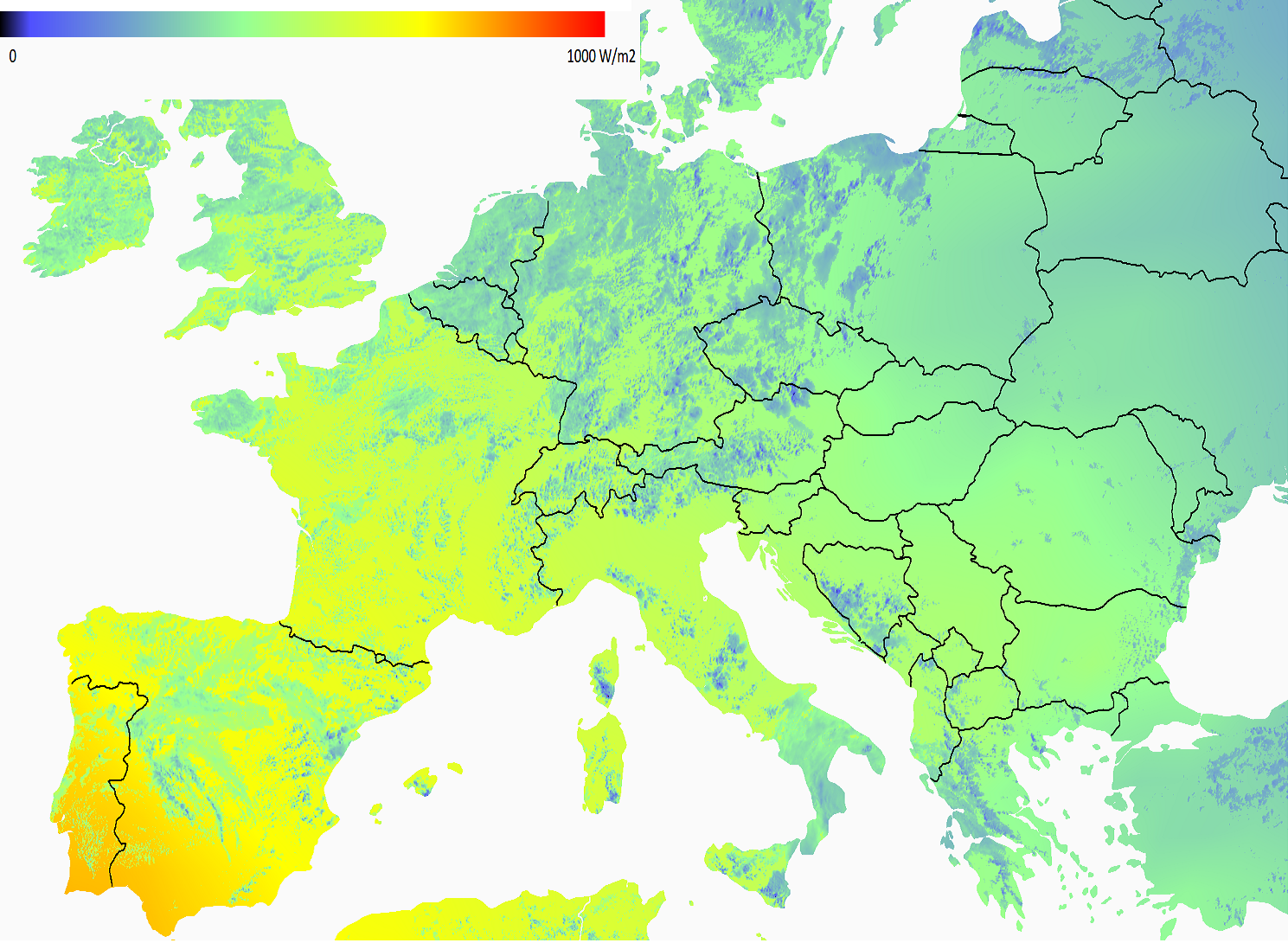}
\caption{Solar irradiance map over Europe (GHI) (20/08/2021 @11:00 CET).}
\label{fig:solar}
\end{figure}

Designing a system that uses solar energy for computing requires numerous decisions concerning energy being generated. Solar energy generation depends on a geographical location and, at the same time, varies throughout the year as solar irradiation varies significantly between seasons. The primary decision to be made is the capacity of the photovoltaic system. Its size depends mainly on the geographical location and the number of cloudy days during the year. Temporary cloudiness, as presented in Fig.\ref{fig:solar} is very dynamic and introduces uncertainty in energy management. The size of the photovoltaic system is designed based on annual averages, where the generated energy surplus is transferred/sold to the grid in the summer. In winter, it is received/purchased from the grid. This mechanism makes it possible to design systems whose annual energy demand is covered by renewable energy. Unfortunately, such a solution involves the cost of transferring/receiving or selling/buying energy to the grid.

\begin{figure}[t]
\centering
\includegraphics[width=0.8\textwidth]{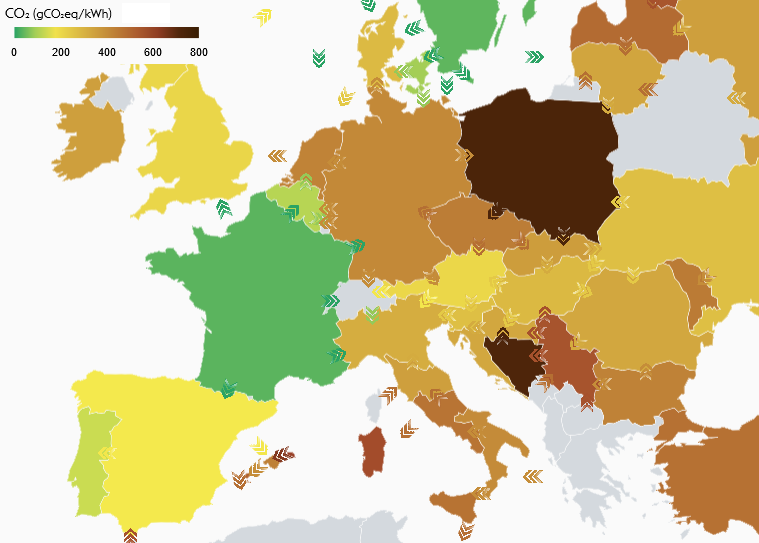}
\caption{Carbon emission map (\textit{source:electricitymap.org})}
\label{fig:electricity}
\end{figure}

Reducing the energy amount being sold/purchased to/from the grid will maximize the utilization of the designed solar energy generating system. Subsequently, data processing on the edge of the network can be optimized to better use solar energy without the need to transfer it to/from the grid. This approach may bring significant savings and better energy management.

The power grid is also characterized by indicators of various types of energy providers. These can be multiple types of wind and water turbines, photovoltaic farms or nuclear power plants. As a result, the electricity use in a particular country is related to the amount of carbon dioxide released into the atmosphere for each kWh of energy used as presented in Fig.\ref{fig:electricity}.

\subsection{Energy-aware management}
\begin{figure}[htbp]
  \centering
  \includegraphics[width=0.8\textwidth]{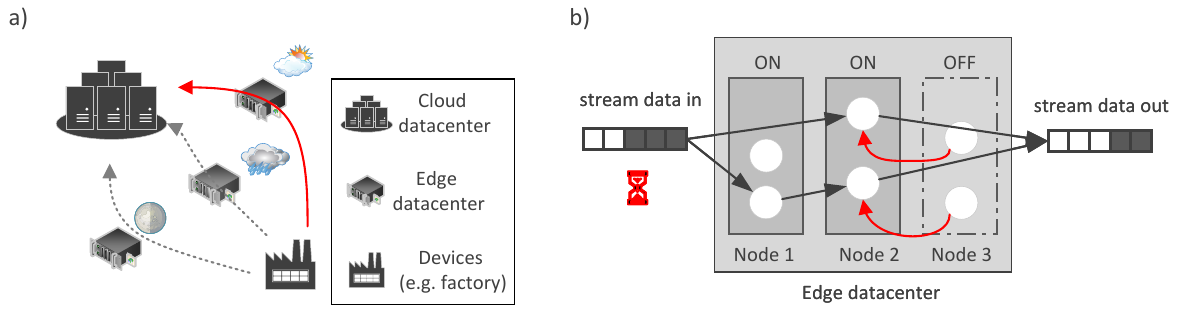}
  \caption{Adaptability mechanisms for stream processing: (a) global (b) local}
  \label{fig:actions}
\end{figure}

The dynamic change in the availability of renewable energy results in the uncertainty of its supply. To ensure the desired properties of the IoT system, it is necessary to adapt the processing of data streams from IoT devices. The adaptability mechanisms applied to the energy-aware computing systems (Fig. \ref{fig:actions}) can be distinguished based on the scale of management:

\begin{itemize}
    \item At the global level, it is possible to switch stream processing between data centres powered from photovoltaic panels, (1)based on the fact that the sun that rises in the east and sets in the west or (2)to the least utilized data centres to preserve energy (Fig. \ref{fig:actions}a).
    \item At the local level - at the edge datacenter (Fig. \ref{fig:actions}b) - it is possible to locally manage data processing in such a way as to adapt to locally available resources and weather conditions. It can be done e.g. by slowing down the streams of the input data or switching off computational nodes thanks to rescheduling processing tasks to other nodes – the processing is not switched to other locations.
\end{itemize}

%We think that if the discussed systems are treated in layered way, the holistic management would be, to some extent, a continuation of the work regarding the holistic management of IT systems. 

Each of the mechanisms can be activated independently, but it can lead to unexpected results. One of the solutions to this problem is to use the mechanism to resolve resource contention issues at runtime, according to user preferences.
In the previous research, we have developed the holistic computing controller (HCC)\cite{SzydloH04}, which operation involves the selection of configuration parameters which – following enactment – specify the expected functional properties of the system. For example, in an HCC-managed system, the system operator identifies the QoS metrics as the cost of operation, data sampling frequency, power saving, system lifetime, and availability. Unfortunately, the central management of the complex architecture used to process data from sensors is ineffective due to the volatility and unpredictability of the environment in which it operates.

%Osmotic computing can be seen as a multi-agent system \cite{}, where agents associated with computing infrastructure elements can communicate and take coordinated management actions. Issues related to the implementation of osmotic agents are discussed in more detail in the next section.

\subsection{Autonomic Computing}
The complexity of multilayer IoT systems and the dynamism resulting from renewable energy sources make the construction of management mechanisms complex. The most promising solution is to use the concept of Autonomic Computing (AC) \cite{ac}, in which the systems have the ability to self-manage. The main idea is that computer systems undertake adaptation actions based on high-level policies given by the administrators. Such systems monitor the system's state and plan actions in response to changing conditions of their operation. In the case of complex, distributed computing systems, autonomic computing can be viewed as a multi-agent system \cite{MAS} where agents manage system components and communicate to undertake coordinated adaptation actions.

%In advanced systems, agents can learn which action to take using machine learning. Reinforcement learning methods can be used by agents deployed in devices to learn autonomously how the actions taken impacted the change of the device state and the obtained reward or penalty. Such knowledge can be then distributed to other osmotic agents to update their adaptation logic. Knowledge exchange regularly should accelerate the learning process and, most importantly, constantly update cooperating agents' adaptation logic during the IoT system runtime fulfilling the processing requirements of the whole system.

\tomasz{The operation context uncertainty of IoT systems, mainly due to the specificity of their operation, causes management decisions to move from the design phase to the execution phase. As a result, responsibility for management actions is shifted from developers to the system itself. The solution for the implementation of agent logic is the usage of machine learning algorithms, which can improve their performance during their work \cite{ml_sas}. However, the promising approach is reinforcement learning\cite{DRL} which principle is based on interacting with the system by performing actions on it and observing how they influence its operation. The constant feedback from the system is especially applicable in systems characterized by high uncertainty.}

In osmotic computing, autonomic agents deployed on the computing nodes can be perceived as a distributed multi-agent system \cite{osmotic_agents} that manages the proper execution of the data flows. The osmotic agents can share knowledge with other agents through the osmosis membrane.

\section{Energy modelling}

\begin{figure}[t]
\centering
\includegraphics[width=0.7\textwidth]{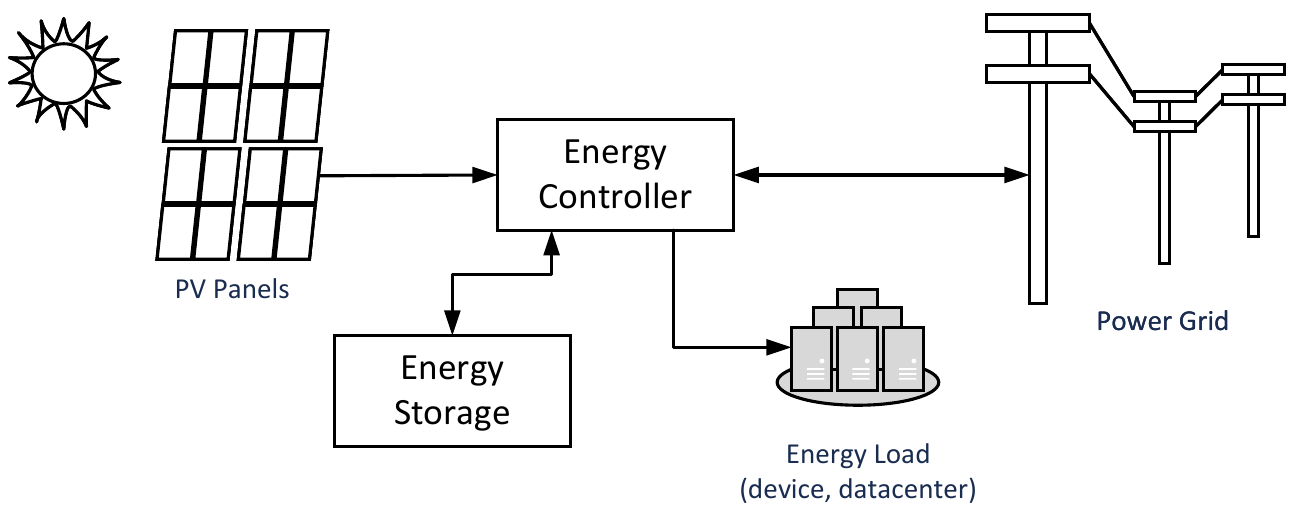}
\caption{Datacenter powered by renewable energy.}
\label{fig:energy_topology}
\end{figure}

In the \textit{IoTSim-Osmosis} extension, we have adopted the topology of electrical connections, including renewable energy sources, energy storage, and the power grid, as shown in Fig.\ref{fig:energy_topology}. The $EnergyController$ module associated with each datacenter is responsible for energy management. Symbols used in the equations are summarized in the Tab.\ref{tab:notations}. Energy controller module is described by a tuple:

\begin{equation}
E_i = (P_i, S_i, B_i, D_i, p_i, e_i^u)
\end{equation}

\noindent where $P_i = (p_i^{cost}, p_i^{low}, p_i^{res})$ stands for power grid and the corresponding parameters. The configuration of the photovoltaic panels is described as $S_i = (s_i^{peak}, s_i^{angles})$. Surplus energy is stored in the battery $B_i = (b_i^{cap}, b_i^{s})$.

The energy consumed by the datacenter $D_i$ is the result of the processing of data streams in it. Several factors are considered when designing renewable energy solutions, including geographical location and average annual energy use. It is then possible to match the size of the photovoltaic installation to the actual energy demand. Usually, the PV installation capacity is taken as a percentage of the annual energy demand. Thus, the average value of energy consumed by the data centre can be estimated according to the formula:

\begin{table}[t]
\centering
\caption{Notations used in the system model.}
\label{tab:notations}
\begin{tabular}{ll}
\hline
Symbol & Meaning                             \\ \hline
$E_i$      & Energy controller at datacenter $i$ \\
$P_i$      & Power grid at datacenter $i$ \\
$S_i$      & Renewable energy source at datacenter $i$ \\
$B_i$      & Battery at datacenter $i$ \\
$D_i$      & Datacenter $i$ \\

$p_i$      & Energy management policy used by energy controller at \\
           & datacenter $i$ \\

$e_i^u$      & Renewable energy utilization percentage at datacenter $i$ \\

$e_i$      & Average power consumption at datacenter $i$ \\
$e_i^{re(t)}$   & Renewable energy available at $D_i$ at a particular time point $t$\\

$s_i^{ann}$   & Annual energy generated by solar panel at datacenter $i$\\

$p_i^{cost}$   & Cost of the kWh of energy in the country where datacenter $i$ \\
              & is located \\
$p_i^{low}$   & Percentage of the low-carbon emission sources in the country \\
              & where datacenter $i$ is located \\

$p_i^{res}$   & Percentage of the RES in the country where datacenter $i$\\
              & is located \\
$t_i^{j}$   & Processing time of the osmotic transaction $j$ on datacenter $i$ \\
\hline
\end{tabular}%
\end{table}

\begin{equation}
e_i = \frac{s_i^{ann}}{e_i^{u}} \cdot \frac{1}{365 \cdot 24}
\end{equation}

Renewable energy obtained from the sun is considered in annual cycles. The surplus of energy obtained in summer balances the energy obtained in the winter months. Therefore, $s_i^{ann}$ is calculated as the total solar energy harvested at datacentre $i$ during a year.
It is assumed that the datacenter consumes energy at a constant level, resulting from the percentage of energy obtained from the PV installation for the year in which the simulation is carried out. Any shortfall in PV energy needed to power the data centre is supplemented by energy from the power grid or battery depending on the provided energy management policy $p_i$.

%The simulator works for data centres that are continuously connected to the power grid, so-called on-grid systems. This means that the surplus electricity generated by the PV plant is virtually stored in the power grid.

Renewable energy consumed directly by the data centre is self-consumed without being sent to the power grid. In sustainable systems, the aim is to increase the self-consumption rate because it reduces the amount of energy flowing through the power grid. In the case of the proposed model, the level of self-consumption is expressed by the formula:

\begin{equation}
e_i^{self(t)} = 
\left\{\begin{matrix}
1 & \quad e_i^{re(t)} > e_j^m\\ 
\frac{e_i^{re(t)}}{e_i} & \quad e_i^{re(t)} \leqslant e_i 
\end{matrix}\right.
\end{equation}

\begin{figure}[ht]
\centering
\includegraphics[width=0.70\textwidth]{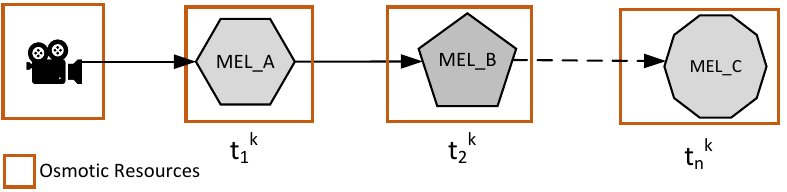}
\caption{Osmotic flow transaction.}
\label{fig:calculation}
\end{figure}

The simulator also incorporates various metrics based on renewable energy use from photovoltaic panels and the power grid. We assumed that the power grid has both a price and a carbon footprint corresponding to the kWh of energy consumed by the data centers. The carbon footprint is mainly driven by the type of energy sources, including renewable, low-carbon and fossil fuel-based. In metric value calculation it is assumed that the osmotic application consists of a series of transactions occurring one after another. The Fig.\ref{fig:calculation} shows a transaction involving the flow of data streams through the different MELs. In particular:

\begin{itemize}
    \item The energy self-consumption metric $M_{self}$ takes into account the total processing time of the data streams by the MEL in datacenters. In addition, temporal information of the amount of renewable energy from photovoltaic panels are used.

\begin{equation}
M_{self} = \sum_{k}^{}\left ( \frac{ \sum_{i}^{} t_i^k e_i^{self(t_k)}  }{\sum_{i}^{}t^{_{i}^{k}}} \right )
\end{equation}
    
    \item The metric $M_{low}$ for the use of low-emission sources considers both the self-usage of energy from photovoltaic panels powering the datacentre and the annual share of such sources when using the power grid.

\begin{equation}
M_{low} = \sum_{k}^{}\left ( \frac{ \sum_{i}^{} t_i^k e_i^{self(t_k)} + \sum_{i}^{} t_i^k p_i^{low} (1-e_i^{self(t_k)})}{\sum_{i}^{}t^{_{i}^{k}}} \right )
\end{equation}
    
\begin{figure}[t]
\centering
\includegraphics[width=0.55\textwidth]{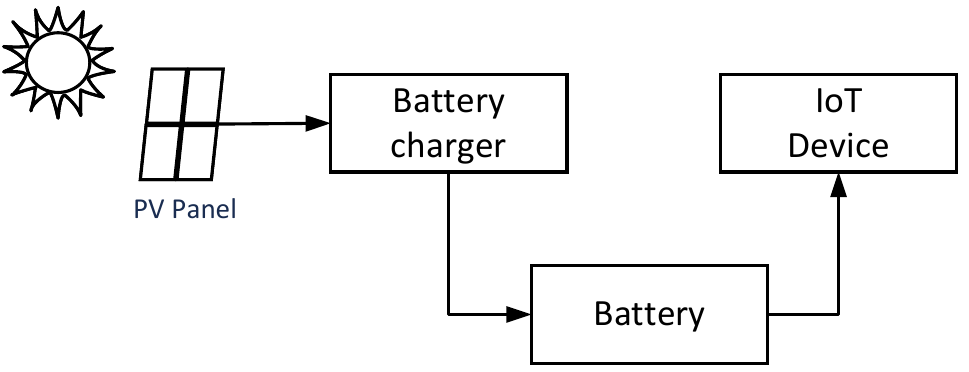}
\caption{IoT device powered by renewable energy.}
\label{fig:energy_device}
\end{figure}    

\end{itemize}

For IoT devices, by default, they are battery-powered. In the case of the discussed extension, it was assumed that they could also be powered by solar energy. The topologies of electrical connections for IoT devices are shown in Fig. \ref{fig:energy_device}. The previously used energy model of the device has been extended with a charging module. Energy from PV panels is used to charge the battery directly, assuming the limitations (1) the battery cannot store more energy than the nominal capacity, (2) the maximum charging current is limited and configurable by the user. The second criterion was introduced because too high charging current reduces the battery's maximum capacity in a longer time horizon.

\section{Design of the simulator}
\begin{figure}[t]
\centering
\includegraphics[width=0.99\textwidth]{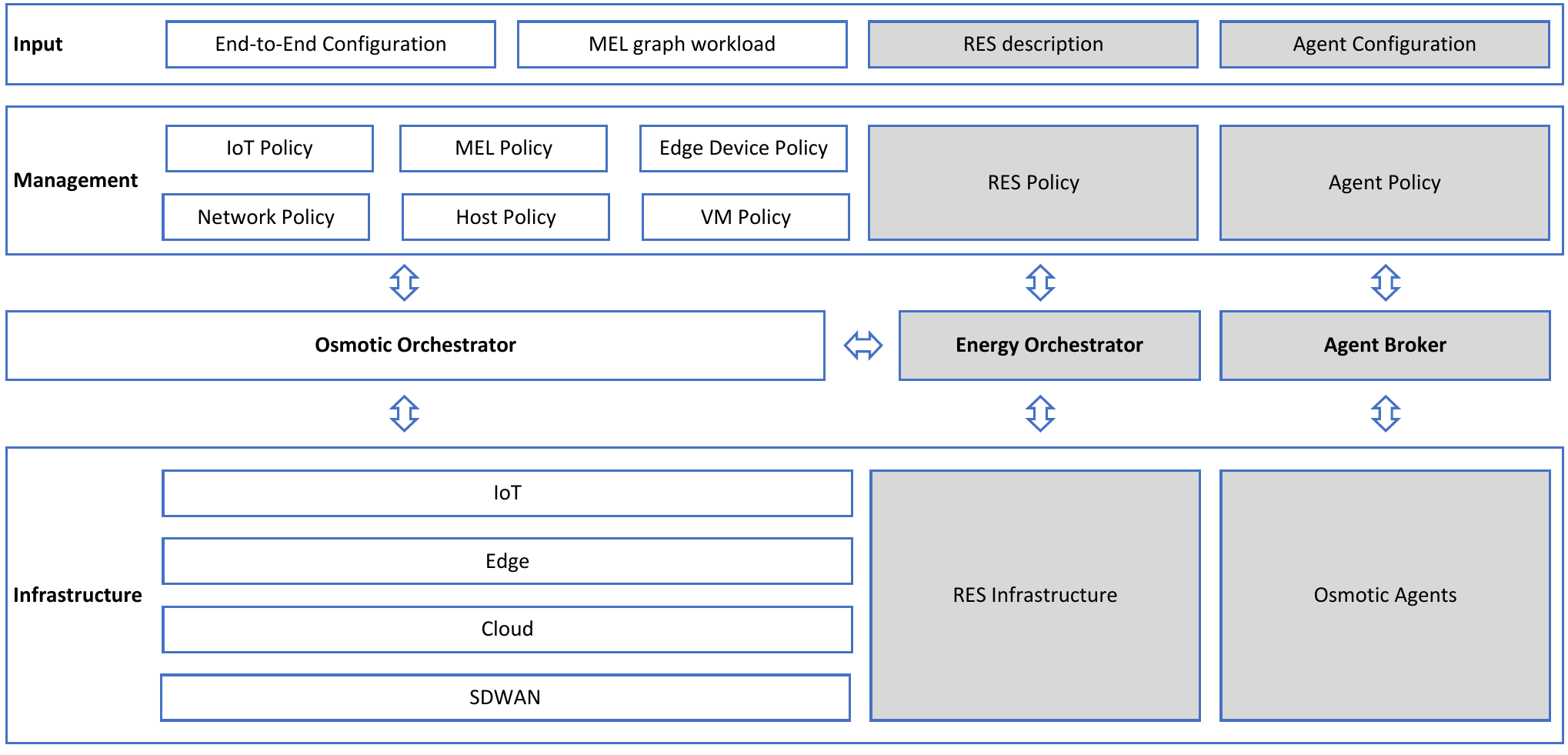}
\caption{Architecture of the simulator.}
\label{fig:architecture}
\end{figure}

The proposed simulator is based on the \textit{IoTSim-Osmosis} and extended by renewable energy modules and the osmotic agent mechanisms enabling research on renewable energy-aware autonomic IoT systems. The high-level architecture of the solution is shown in Figure \ref{fig:architecture}. The discussed solution consists of two modules - the first one responsible for modelling renewable energy sources and the second one implementing the concept of osmotic agents.

\subsection{Renewable Energy Sources module}
\begin{figure}[t]
\centering
\includegraphics[width=0.65\textwidth]{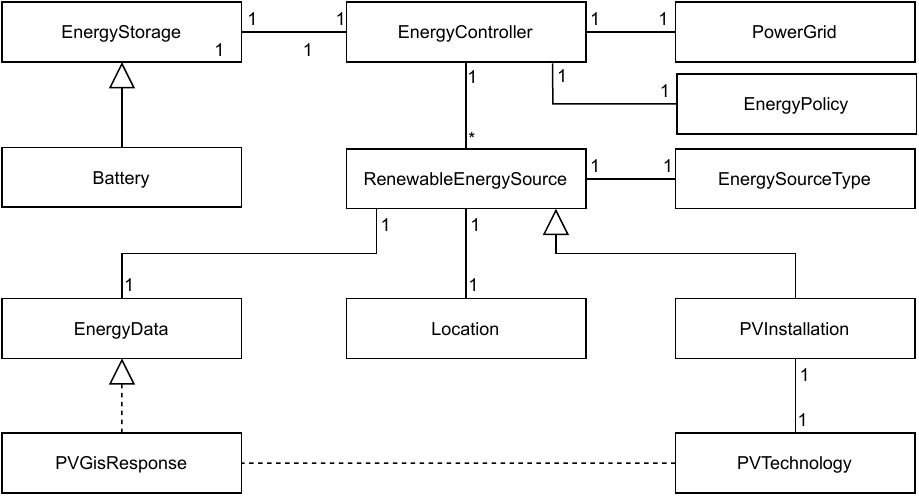}
\caption{Conceptual diagram of the renewable energy module.}
\label{fig:res_classes}
\end{figure}

There are a few primary sources of green energy as solar, wind, water. However, in the simulator, the focus is on solar energy with the future possibility of extensions to others energy sources. The power supply scheme of each datacentre is shown in the Fig.\ref{fig:energy_topology}, while the details are presented in Fig. \ref{fig:res_classes}.
%We made a package for PVGIS data, modelling energy sources, storage for excess of energy, and policies responsible for dealing with produced energy and model for our PV panels and power grids. This model also leaves place for other sources of renewable energy.

\subsubsection{Energy Sources}

The module allows simulating the operation of a photovoltaic installation connected to the selected datacentre. The configuration includes parameters such as geographic location, power of the installation and an indicator informing about the percentage share of the produced energy in the annual energy consumption of the datacenter.

%As it stands for the name of the package, the main component is the energy source. It represents a green energy source like solar panels. It contains information about data from PVGiS, which tell us the amount of generated green energy at a particular point of time for that PV (photovoltaics) installation. For that purpose, we have created the special parsers for PVGiS and an additional model package where we implement basic classes responsible for holding parsed information. Also, the energy source stores data about its location, type and unique ID by which we can easily link it with the right datacenter.

Each energy source corresponds to one PV installation. However, some datacenter may require energy that could not be provided by only one installation. Furthermore, since the solar energy generated by PV panels is hesitant and depends on several factors, a few of them are required to specify before the simulation, such as angle of PV panels, temperature power loss and peak power.

Historical data are downloaded from the Photovoltaic Geographical Information System (PVGIS) portal\footnote{https://ec.europa.eu/jrc/en/pvgis}. It provides data about solar radiation and PV potential based on satellite image analysis for the 2005-2016 years and can give more detailed monthly, daily, and hourly intervals.

%That is why we provide for users additional configuration files where all those details need to be specified.

\subsubsection{Energy Storage}
Energy storage devices are used to store surplus energy produced by PV panels and then use it later in the day. They can provide energy when needed, especially in the situation of reduced generation from renewable sources, if it has been previously stored. If the energy storage is fully charged, the surplus energy is transferred to the power grid. They are particularly important for reducing the characteristic increase in energy generated at noon by distributed photovoltaic installations and the significant demand after sunset. 

Energy storage can be considered as a solution that balance energy in a daily time horizon. Several technological solutions are available\cite{storage} such as rechargeable batteries, pumped hydroelectric store, compressed air storage, or flywheels. However, in the extension, we are focusing on rechargeable battery technology. Therefore, battery parameters, i.e. its total capacity and current charge level, are determined at the beginning of the simulation.

%We cannot overcome the nature and changeability of sunshine in a day, wind speed, water flows etc. Such uncertainty makes our RES periodical. Hence datacenters can not always rely on them. It means that sometimes, the energy generated by RES may not meet the energy consumption of DC. On the other side, the energy produced by RES could be exceeded and will be lost if datacenter do not consume a large amount of energy while accomplishing jobs. For that purpose, we introduce energy storage to be charged when RES is producing too much energy and, on the other hand, to let DC use that energy when it is needed.

\subsubsection{Power grid}
There could be a situation when the demand for energy consumption of DC is much higher than the amount that any renewable energy source can produce. In such case, the power grid supplies energy to DC by being switched on by the controller, which checks whether is there enough energy coming from, e.g. photovoltaic panels. Therefore, it is important to specify the country where the installation is located in the RES configuration file and the price for the energy taken from the power grid.

\subsubsection{Energy Policy}
The user-selectable policy performs management of the energy controller. It defines the way how the controller should take care of the energy consumption and the situation when demands of energy differ from the amount coming from renewable installation.

In the module there are provided three policies to cover real-world cases of energy management, which can be extended by the users developing their own strategies. Names of the available policies are self-explanatory and are:
\begin{itemize}
    \item $GridOnlyPolicy$ - datacenter is connected to the power grid only without energy storage device and the renewable energy source,
    \item $OnGridpolicy$ - datacenter uses energy from the power grid as well as from the renewable energy source, surplus of the energy is transferred to the power grid,
    \item $OnGridEnergyStoragePolicy$ - similarly as in the $OnGridpolicy$ but the datacenter is also equipped with energy storage device.
\end{itemize}
 Before the simulation, there is a requirement to chose defined policies in the configuration file.

%\begin{itemize}
%    \item PVGIS - Photovoltaic Geographical Information System, it provides data %about solar radiation and PV potential. It relates to the 2005-2016 years and can %give more detailed monthly, daily, and hourly intervals.
%\end{itemize}

\subsection{Osmotic Agents module}
\begin{figure}[t]
\centering
\includegraphics[width=0.4\textwidth]{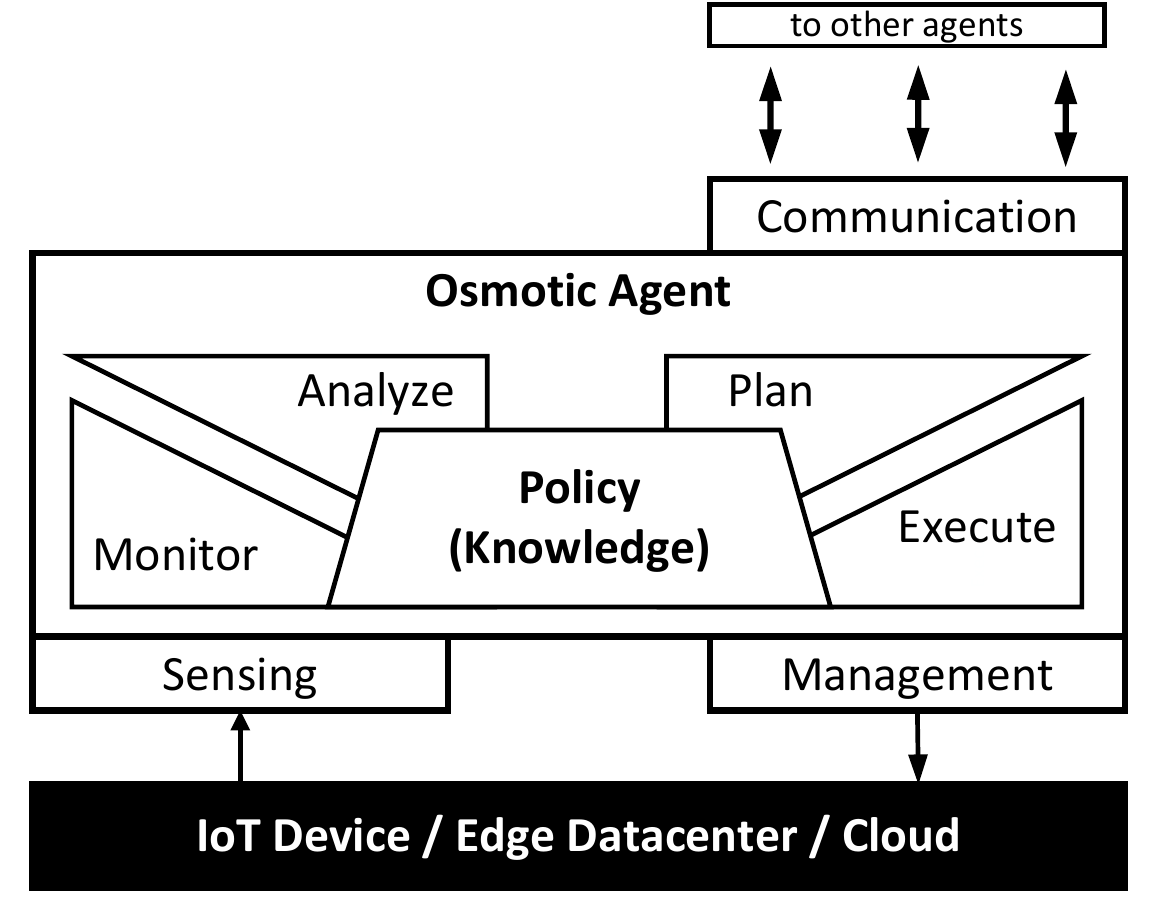}
\caption{Autonomic Osmotic Agent.}
\label{fig:agent}
\end{figure}

In the simulator, we have implemented autonomous agents, as shown in Figure \ref{fig:agent}. Agents are assigned to IoT devices as well as edge and cloud datacenters. Agents implement the autonomic computing MAPE loop\cite{mape} - \textit{Monitor}, \textit{Analyze}, \textit{Plan} and \textit{Execute}. \tomasz{Osmotic agent structures the control and management loop of the system, distinguishing the stages of monitoring, analyzing, planning and executing actions on the system. In the simplest case, the operating logic of an osmotic agent can be implemented in the form of a rule-based manner, where a set of system adaptation rules are prepared during the development phase and are not changed at runtime.}

\tomasz{In the case of reinforcement learning and the so-called reinforcement agent, environment observation occurs during the monitoring phase of the osmotic agent, where knowledge about the environment is obtained using dedicated sensors. Then the data is analyzed, and in the planning phase, appropriate management actions are selected by the RL agent. Finally, they are enforced in the execute phase using dedicated effectors.}

Agents lifecycle is controlled by the \textit{AgentBroker} component. It contains the topology of the infrastructure, thus is responsible for message passing. Moreover, it controls MAPE execution based on the simulator's internal timer. Interface-based architecture facilitates the end-user to implement logic for the agents.

\subsubsection{Cooperation models}
\begin{figure}[t]
\centering
\includegraphics[width=0.99\textwidth]{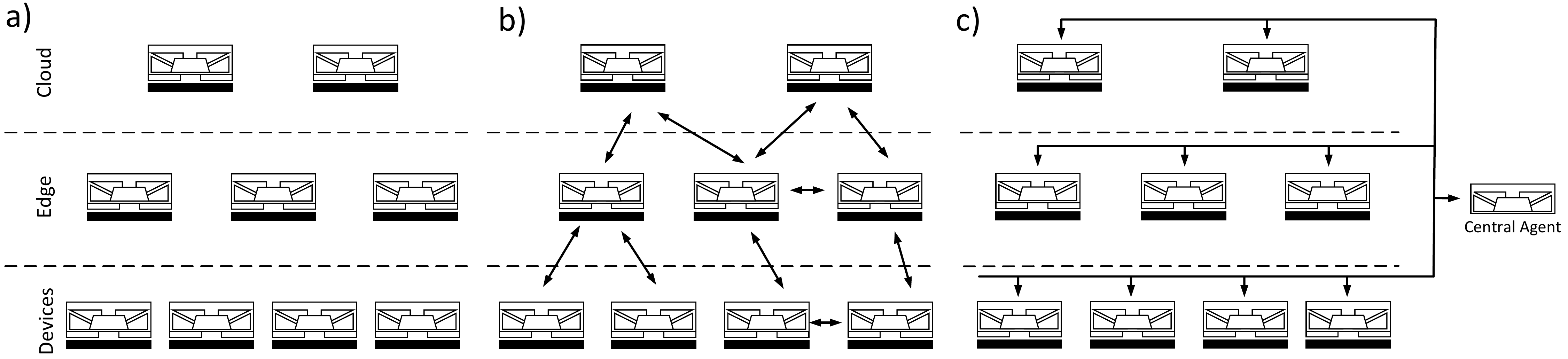}
\caption{Supported communication patters between agents: a)independent agents/learners with independent policies, b)decentralized agents exchanging information with the neighbouring agents but acting independently, c) centralized training with decentralized execution.}
\label{fig:cooperation}
\end{figure}

The osmotic agents can constitute the distributed multi agent environment in which they can cooperate by exchanging messages. The provided module enables the implementation of three scenarios of cooperation between agents (Fig.\ref{fig:cooperation}):
\begin{itemize}
    \item \textit{independent agents} - each agent manages a device or a data centre independently of other agents. Agents do not exchange messages.
    \item \textit{communicating agents} - agents can exchange messages, but they make management decisions and actions independently.
    \item \textit{central agent} - the central agent receives information from all other agents and makes decisions regarding each device that the agents enforce. Local agents only perform the operations requested by the central agent.
\end{itemize}

\begin{figure}[t]
\centering
\includegraphics[width=0.7\textwidth]{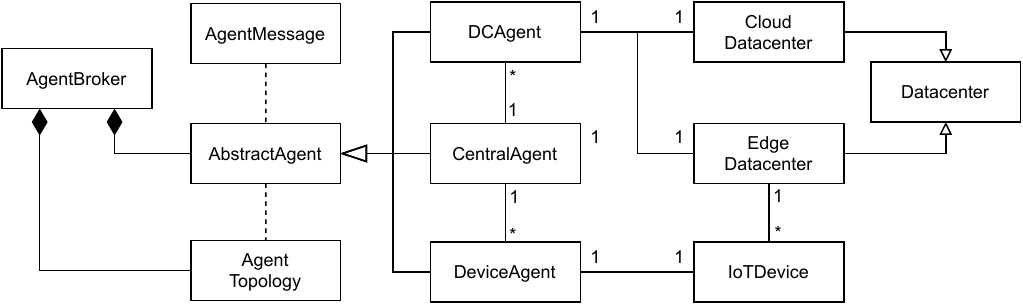}
\caption{Osmotic Agents structure.}
\label{fig:agents_architecture}
\end{figure}

%\begin{figure}[t]
%\centering
%\includegraphics[width=0.99\textwidth]{images/agent_class_v01.png}
%\caption{Osmotic agent class.}
%\label{fig:architecture}
%\end{figure}

\subsection{Adaptability actions}
\begin{figure}[h]
\centering
\includegraphics[width=0.35\textwidth]{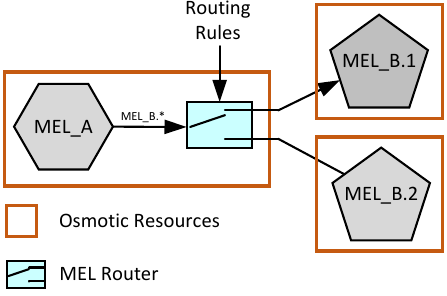}
\caption{MEL routing concept.}
\label{fig:routing_principle}
\end{figure}

Because of the fact that agents are associated with the datacenters, they can apply adaptation action in the execute phase of its internal control loop. It may include modification of all the parameters related to the datacenter itself, such as processing power and a number of available processing nodes. Nevertheless, to enable processing flow adaptability, the concept of adaptive routing was introduced as presented in Fig.\ref{fig:routing_principle}. In that concept, \textit{MEL\_B.*} is a stateless abstract \textit{MEL\_B} implementing processing functionality, but not pointing to the particular instance.

The \textit{MEL\_B.1} and \textit{MEL\_B.2} are instances of abstract \textit{MEL\_B} and are located in different edge datacenters. The selection of the desired functionality implementation is made in the data stream routing process,  controlled by appropriate routing rules. By default, matching instances are selected based on the \textit{Round-Robin} policy if there are no specific \textit{MEL} routing rules provided.

%In configuration we created two MELs -- \textit{MEL.1} and \textit{MEL.2}. \textit{MEL.} is inferred to \textit{MEL.1} or \textit{MEL.2}, and by default, it is driven by the Round-Robin policy.

\section{Evaluation of IoTSim-Osmosis-RES}
The section presents the results of evaluating the proposed framework. The results of five algorithms for managing the operation of an example IoT system and the obtained results are presented.

\subsection{Infrastructure}
The framework evaluation is carried out for the IoT system that processes data from IoT devices. The system processes video streams from cameras installed in the smart city. Data streams are pre-processed in edge data centres and then processed in the cloud.

The analysis is performed for two edge data centres located in Berlin and Paris and one cloud datacenter located in Ireland. All of them are powered by photovoltaic panels. Tab.\ref{tab:infra} contains the details of the hardware infrastructure. the smart cameras are installed in Berlin, powered by photovoltaic panels and equipped with rechargeable batteries. Details of the devices used in the evaluation are presented in Tab. \ref{tab:dev_exp}.

\begin{figure}[h]
\centering
\includegraphics[width=0.7\textwidth]{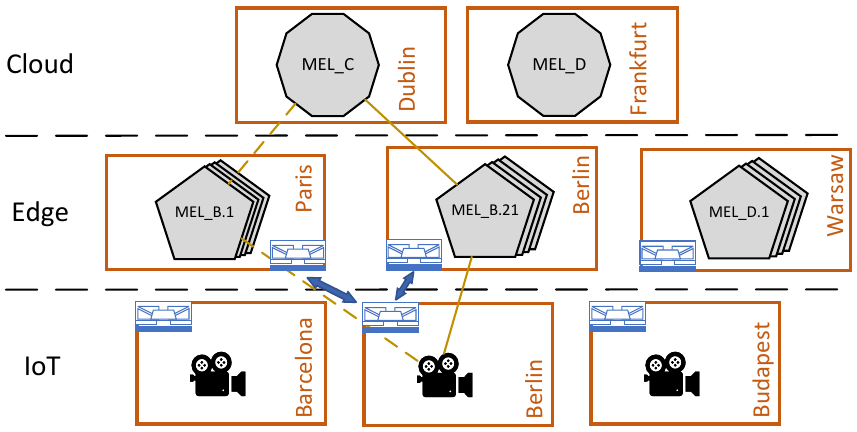}
\caption{Osmotic Agents structure.}
\label{fig:case_all}
\end{figure}

\begin{table}[]
\caption{Notations used in the system model.}
\label{tab:infra}
\centering
\begin{tabular}{|l|l|l|l|}
\cline{1-4}
                 \textbf{Parameter} & \textbf{Edge1}  & \textbf{Edge2} & \textbf{Cloud}  \\ \hline
\multicolumn{1}{|l|}{Location}  & Berlin & Paris & Dublin \\ \hline
\multicolumn{1}{|l|}{Latitude} & 52.52  & 48.8  & 53.35  \\ \hline
\multicolumn{1}{|l|}{Longitude} & 13.40  &  2.30 & -6.30   \\ \hline
\multicolumn{1}{|l|}{Type}      & on-grid&on-grid&on-grid \\ \hline
\multicolumn{1}{|l|}{Battery}   & no     & no    & no     \\ \hline
\multicolumn{1}{|l|}{RES Energy Utilization} & 60\%   & 60\%    &  40\%    \\ \hline
\end{tabular}%
\end{table}

\begin{table}[]
\caption{IoT device battery parameters.}
\label{tab:dev_exp}
\centering
\begin{tabular}{|c|c|c|c|c|}
\hline
Device Type   & Battery Capacity & Initial  Energy & Battery Voltage & Solar Panel \\ \hline
Smart  camera & 3000mAh          & 2000mAh         & 3.7V            & 10W         \\ \hline
\end{tabular}%
\end{table}

\subsection{Osmotic Agents}
Each device used in the evaluation has assigned \textit{Osmotic Agent}($DcAgent$ for Datacenter, $DeviceAgent$ for IoTDevice). Its job is to decide whether data should be computed locally or in another datacenter and thus, reroute the processing flow to another location. In the experiments we consider five algorithms for \textit{Osmotic Agents}:
\begin{itemize}
    \item \textbf{(ALG1/2)} Static selection of edge datacenter processing locations. In our case, these are Berlin(1) or Paris(2).
    \item \textbf{(ALG3)} Selection based on the \textit{Round-Robin} algorithm - data is processed alternately between edge datacenter where MELs required by the application are available.
    \item \textbf{(ALG4)} Adaptive edge datacenter selection algorithm based on the amount of renewable energy available on the edge datacenter, and in case several datacenters are fully powered by renewable energy at any given time, the closer datacenter is selected. During nighttime hours, the edge datacenter that is powered by a power grid with a higher percentage of low-carbon sources is selected. The algorithms for \textit{Osmotic Agents} are presented in listings \ref{alg:dc_agent} and \ref{alg:device_agent}.
    \item \textbf{(ALG5)} Analogously as before, but when none of the computing centres is powered by solar energy from the installed PV panels, the geographically closest one is selected.
\end{itemize}

\begin{algorithm} [!h]
  \caption{DC Agent} \label{alg:dc_agent}
  \begin{algorithmic}[1] 
    \Statex
    \Function{Monitor}{ }
        \State{ $r\gets$ solar irradiation value}
        \State{ $S\gets$ {available MELs in DC}}
        
    \EndFunction
    \Statex
    \Function{Analyze}{ }
        \State{$m\gets \varnothing$}
        \State{$m.dest \gets$ neighbouring IoT devices}
        \State{$m.content \gets (r,S)$}
        \State{$publish(m)$}
    \EndFunction
    \Statex    
    \Function{Plan}{$M$}
        \State{ $nothing$}
    \EndFunction
    \Statex
    \Function{Execute}{ }
        \State{ $nothing$}
    \EndFunction
  \end{algorithmic}
\end{algorithm}

\begin{algorithm} [!h]
  \caption{Device Agent} \label{alg:device_agent}
  \begin{algorithmic}[1] 
    \Statex
    \Function{Monitor}{ }
        \State{ $nothing$}
    \EndFunction
    \Statex
    \Function{Analyze}{ }
        \State{ $nothing$}
    \EndFunction
    \Statex    
    \Function{Plan}{$M$}
        \State{$Q\{\} \gets \varnothing$}
        \State{$V\{\} \gets \varnothing$}
        \For{each $m_i \in M$}
            \State{$(r,S) \gets m_i.content$}
            \For{each $s_j \in S$}
                \State{$s' \gets abstract(s_i)$}
                
                \If{$s' \not\in Q.keyset$}
                    \State{$V\{s'\} \gets r$}
                    \State{$Q\{s'\} \gets s_j$}
                \ElsIf{$V\{s'\} < r$}
                    \State{$V\{s'\} \gets r$}
                    \State{$Q\{s'\} \gets s_j$}
                \EndIf
            \EndFor
        \EndFor
    \EndFunction
    \Statex
    \Function{Execute}{ }
        \For{each $s' \in Q.keyset$}
            \State{$routing\_add(s',Q\{s'\})$}
        \EndFor
    \EndFunction
  \end{algorithmic}
\end{algorithm}

Algorithms ALG4 and ALG5 use \textit{Osmotic Agents} to manage the processing of data streams from IoT devices. In the MAPE loop, agents are processing received messages from other agents and modify routing tables in assigned devices. After these steps, data flow can change destination, which implies flow processing being moved to the desired place. In the algorithm ALG5, the geographically closest computing centre selection is based on the haversine algorithm, which determines the distance of a great circle between two points on the globe.

\tomasz{The discussed algorithms run simultaneously on edge datacenter and devices. In the case of edge datacenter, the algorithm sends a message containing the deployed MELs list to the available IoT devices. Therefore, the algorithm's complexity is linear to the number of available MELs on a particular edge and the number of IoT devices. In the case of devices, the algorithm processes messages from each edge datacenter, and then for each MEL from the messages, the one that meets the appropriate criteria is selected. Therefore, the algorithm's complexity is linear to the number of edge datacenters and the total number of MELs available in the system.}

\begin{figure}[h]
\centering
\includegraphics[width=0.99\textwidth]{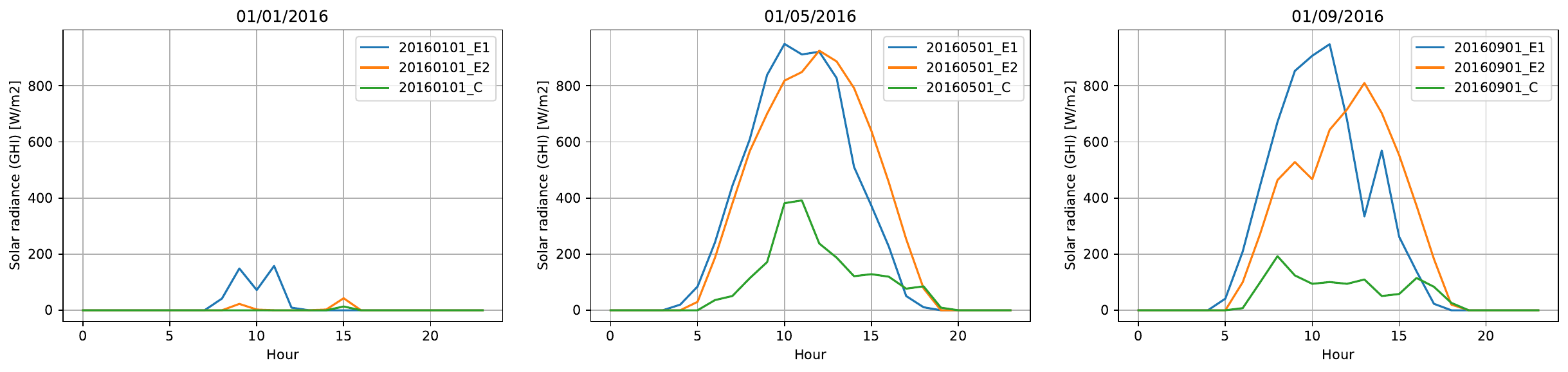}
\caption{Solar radiation during the experiments.}
\label{fig:ghi}
\end{figure}

\begin{figure}[h]
\centering
\includegraphics[width=0.99\textwidth]{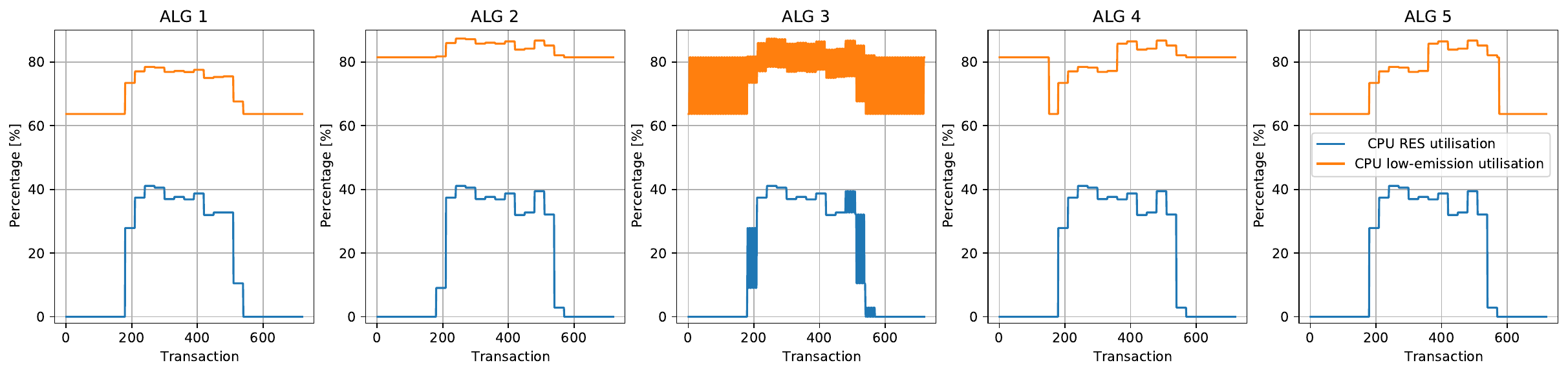}
\caption{Usage of the RES and low emission sources during the experiment.}
\label{fig:alg}
\end{figure}

\begin{figure}[h]
\centering
\includegraphics[width=0.99\textwidth]{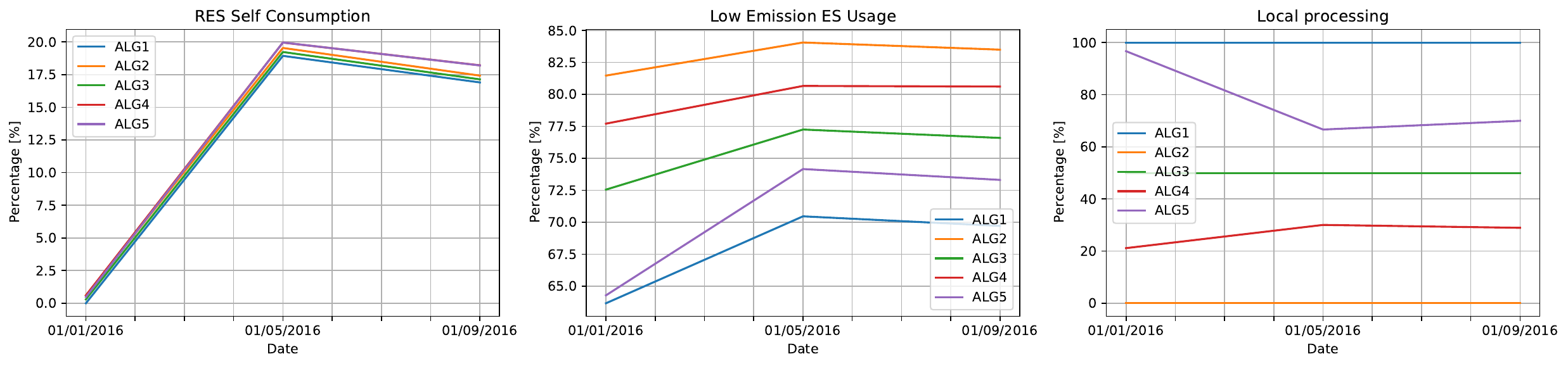}
\caption{Comparison of the algorithms used for adaptive management of processing.}
\label{fig:metrics}
\end{figure}

\begin{figure}[h]
\centering
\includegraphics[width=0.33\textwidth]{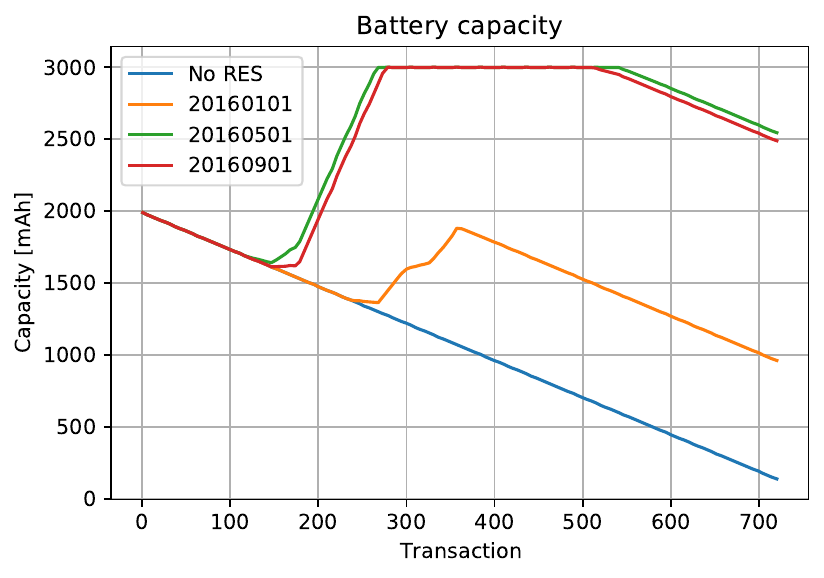}
\caption{IoT device battery capacity during the experiment.}
\label{fig:battery}
\end{figure}

\subsection{Results}
Fig. \ref{fig:ghi} shows the level of solar radiation for selected days at the location of the analyzed edge and cloud computing centres. Changes in the level of solar radiation during the day are the result of cloud cover movement over the sky. The lowest level of solar radiation was noticed in January.

Fig. \ref{fig:metrics} shows a comparison of the analyzed algorithms in the context of different evaluation metrics. These are (1)self-consumption of renewable energy sources, (2)usage of low-emission energy sources, and (3)processing at the nearest edge datacenter for IoT devices. Each algorithm is evaluated using three metrics. For the first metric, related to the self-consumption of green energy, the best results are achieved with the ALG4 and ALG5 algorithms. This is because the logic of osmotic agents adapts the data flow by selecting the edge datacenter with the highest share of green energy to power them. It is also interesting that despite obtaining comparable results for the self-consumption of renewable energy, the algorithms achieved different values for the other metrics. ALG4 prefers data centres connected to the power grid with a higher percentage of low-carbon energy sources, while ALG5 prefers the closest datacenters to the smart cameras. In the case of the second metric, the best results were obtained for the ALG2 algorithm, for which the edge datacenter was statically selected in a country with more nuclear power plants (France). In the case of the last metric, the best results were achieved for ALG1 - the data was always processed in the nearest computing centre (Berlin). 

Fig. \ref{fig:battery} shows the charge level of batteries used in IoT devices for three selected days of the year. During the experiments, it was assumed that the batteries were charged to 66\% at midnight. The energy stored in the battery allows the device to operate for 24 hours - the entire experiment. In the case of the operation of the device in the summer period, the amount of recharged energy from solar panels is greater than the device consumption. This means that the device can work for the next day without interruption - the battery level at the end of the day is higher than at the beginning. However, in the case of winter days and heavy cloud cover, the device's battery may be not recharged during the day. Therefore, in real-world scenario, the solution would be to use the backup power supply from the power grid, oversize batteries and photovoltaic panels, or introduce algorithms for adaptive management of the device's operation.

To sum up, the optimal management of osmotic computing is a multi-criteria optimization problem, and various adaptation goals are achieved depending on the selected algorithm. It is also possible to change the adaptation algorithm during the system operation, resulting from a change in the context of the system operation, e.g. a natural disaster.

%The comparison of RES utilization is presented on figures \ref{fig:alg} and \ref{fig:metrics}.

%points the benefits of both - agent and RES systems. As you can see on the figure 16 the total RES utilization for workload CPU processing is around 18.5\%, and for figure 17 is 20.5\%, so it’s 2\% advantage when using agents which helps to use more renewable energy produces by installations in different regions (datacenters). It could be possible by changing the place (datacenters) where the computation happens based on the information about energy production of renewable sources.

\section{Related work}
There are some existing frameworks for modelling and simulating IoT environments and evaluate resource management.
CloudSim \cite{cloudsim} is a toolkit for modelling and simulation of cloud computing environments and evaluation of resource provisioning algorithms. It provides some basic energy-conscious abstractions and resource management techniques to evaluate energy usage. \Fawzy{EdgeCloudSim \cite{sonmez2018edgecloudsim} is based on CloudSim and is designed to handle edge computing's computational and network requirements. EdgeCloudSim's architecture allows a variety of models for studying features of edge computing, such as the edge server model, the mobility model, and the network link model. EdgeFogCloud \cite{mohan2016edge} support different network and energy models. In addition, it can help in founding a resource network and task scheduling by formulating the configuration parameters.} iCanCloud's \cite{icancaloud} main target is to provide the user information about the cost, power, energy of running IoT applications on configured machines. 
GreenCloud \cite{greencloud} simulates an energy-aware cloud environment in order to observe energy utilization consumed by datacenters, also concerning their internal components and techniques to reduce power consumption. 
DCSim \cite{dcsim} is capable of changing VMs state in order to conserve power. CloudSimSDN \cite{cloudsimsdn} extends CloudSim in enabling it to measure power consumption of network devices and minimize it. iFogSim \cite{ifogsim} provides basic resource management policies - \textit{cloud-only} and \textit{edge-ward} in order to measure their impact on various parameters including energy consumption. \Fawzy{MyiFogSim \cite{lopes2017myifogsim} is an extension of iFogSim that focuses on failure simulation, network configuration, and providing virtual machine migration policy support to mobile clients.}
IoTSim-Edge \cite{iotsimedge} considers devices powered by batteries. 
IotSim-Osmosis \cite{iotsimosmosis} is based on CloudSim and IoTSim-Edge, inheriting all their assumptions and properties, including power management awareness, but extending their ability to model SDN-Networks and Osmosis paradigm.
In summary, there are many frameworks that consider energy management, variety of power sources and network infrastructure. However, none of them are able to simulate floating weather conditions, renewable energy sources or provide easily extendable system that enables researches to specify their own VM and power management policies. These features are nonetheless included in proposed simulator.

\begin{table}[]
\centering
\resizebox{\textwidth}{!}{%
\begin{tabular}{|l|l|l|l|l|l|l|l|l|l|}
\hline
\multirow{2}{*}{Simulator} & \multicolumn{9}{c|}{Feature}  \\ \cline{2-10} 
                           & \begin{tabular}{@{}c@{}}Cloud \\ Processing\end{tabular} & \begin{tabular}{@{}c@{}}SDN \\ Support\end{tabular} & \begin{tabular}{@{}c@{}}SD-WAN \\ Support\end{tabular} & \begin{tabular}{@{}c@{}}Network \\ Communication\end{tabular} & \begin{tabular}{@{}c@{}}Edge \\ Processing\end{tabular} & \begin{tabular}{@{}c@{}}IOT \\ Devices\end{tabular} & \begin{tabular}{@{}c@{}}Power \\ Management\end{tabular} & \begin{tabular}{@{}c@{}}Agents \\ System\end{tabular} & RES \\ \hline
CloudSim \cite{cloudsim}                  & x &   &   &   &   &   & &   &   \\ \hline
EdgeCloudSim \cite{sonmez2018edgecloudsim}         &  &  &  & x & x & x &  &   &   \\ \hline
EdgeFogCloud \cite{mohan2016edge}                  &  &  &  & x & x & x &  &   &   \\ \hline
iCanCloud \cite{icancaloud}               & x &   &   & x &   &   &x&   &   \\ \hline
GreenCloud \cite{greencloud}              & x &   &   & x &   &   &x&   &   \\ \hline
DCSim \cite{dcsim}                        & x &   &   & x &   &   &x&   &   \\ \hline
CloudSimSDN \cite{cloudsimsdn}            & x & x &   & x &   &   &x&   &   \\ \hline
iFogSim \cite{ifogsim}                    & x &   &   & x & x & x & &   &   \\ \hline
MyiFogSim \cite{lopes2017myifogsim}       &   &   &   & x & x & x &  &   &   \\ \hline
SimIoT \cite{simiot}                      & x &   &   &   &   & x & &   &   \\ \hline
IotSim-Edge \cite{iotsimedge}             &   &   &   &   & x & x & &   &   \\ \hline
IotSim-Osmosis \cite{iotsimosmosis}       & x & x & x & x & x & x & &   &   \\ \hline
proposed IotSim-OsmosisRES & x & x & x & x & x & x &x& x & x \\ \hline
\end{tabular}%
}
\caption{ Related work summary}
\end{table}

%\subsection{Cloud processing simulators}
IoT systems based on the Osmotic Computing concept also include data processing in the computational clouds. Among the analyzed simulators, CloudSim is one of the most willingly used solutions. It allows for the evaluation and tests management strategies which can improve the performance of cloud infrastructures. This is possible by modelling different components typical for cloud systems like virtual machines or cloud data centres.  The iCanCloud can also use different cloud brokering strategies and observe the efficiency of more resources thanks to more extended output. GreenCloud is very similar to the iCanCloud, but it extends to network cloud infrastructure simulation. DCSim provides dynamic resource management in the cloud to effectively evaluate energy-aware algorithms. It also enables the modeling and simulating of cooling systems. CloudSimSDN, as its name suggests, is an extension of CloudSim and supports Software Defined Networks. It simulates the utilization of hosts, networks and response time of requests in cloud data centres. Finally, the IotSim-Edge is mostly focused on low power communication technologies and protocols for IoT devices and edge processing.

%\subsection{Power management simulators}
To sum up, none of the discussed solutions can model the variety of green energy sources. Instead, they consider the energy as an already existing entity, whose only relevant parameters are its cost and consumption, without considering other metrics, such as self-consumption of renewable energy. Similarly, none of the discussed frameworks implements the mechanism of cooperating agents used for system management.

\section{Conclusions and future work}
The paper presents extensions to the $IoT-Sim Osmosis$ simulator providing functionality for modelling renewable energy sources for powering devices and \textit{Osmotic Agents} mechanism enabling evaluation of autonomic computing algorithms for \textit{Osmotic Computing}. The results show that the choice of the adaptation algorithm affects the achievable performance of the IoT system and, can be selected to fulfill system requirements. 

%Furthermore, the experiment shows that this can lead to the system using mainly low-carbon energy sources or adapting its operation to maximise the direct use of energy from photovoltaic panels.

In the described example, the logic of the agents is predefined at the system design state. As further work, we plan to investigate reinforcement learning algorithms so that individual devices can learn how their decisions affect the system. Reinforcement learning methods can be used by agents deployed in devices to learn autonomously how the actions taken impacted the change of the device state and the obtained reward or penalty. Such knowledge can be then distributed to other osmotic agents to update their adaptation logic. Knowledge exchange regularly should accelerate the learning process and, most importantly, constantly update cooperating agents' adaptation logic during the IoT system runtime fulfilling the processing requirements of the whole system. \Fawzy{It would be desirable to develop a dynamic data streams concept using reinforcement learning to change the data transfer rate of the IoT devices depending on the available energy and to adapt to the projected amount of energy that can be obtained from renewable energy sources}.

\section*{Acknowledgments}
The research presented in this paper was supported by the National Science Centre, Poland under Grant No. MINIATURA/2021/05/X/ST6/00414. This research was funded by EPSRC project, Sustainable urban power supply through intelligent control and enhanced restoration of AC/DC networks, EPSRC-NSFC Call in Sustainable Power Supply, 2019-2022, EP/T021985/1.

\bibliography{cas-refs}

\end{document}